\documentclass[aps, pra, onecolumn,notitlepage]{revtex4}
\usepackage{amsmath}
\usepackage{amsfonts}
\usepackage{amssymb}
\usepackage{graphicx}
\usepackage[sort&compress]{natbib}
\newcommand{\etal}{{\it et al}. }
\begin{document}
\begin{abstract}
We calculate the Kadowaki-Woods ratio (KWR) in Fermi liquids with arbitrary band structures. We find that, contrary to the single band case, the ratio is not generally independent of the effects of electronic correlations (universal). This is very surprising given the experimental findings of a near universal KWR in many multiband metals. We identify a limit where the universality of the ratio, which has been observed experimentally in many strongly correlated electron systems, is recovered. We discuss the KWR in Dirac semimetals in two and three dimensions. In the two-dimensional case we also generalize the KWR to account for the logarithmic factor  in the self-energy. In both cases we find that the KWR is independent of correlations, but strongly dependent on the doping of the system: for massless fermions the KWR is proportional to the inverse square of the carrier density, whereas the KWR for systems with massive quasiparticles is proportional to the inverse of the carrier density.  
\end{abstract}
\title{When is the Kadowaki-Woods ratio universal?}

\author{D. C. Cavanagh}
\author{A. C. Jacko}
\author{B. J. Powell}
\affiliation{School of Mathematics and Physics, The University of Queensland, QLD 4072, Australia}
\maketitle

\section{Introduction}

Fermi liquid theory describes the low temperature behavior of the vast majority of metals extremely well  \cite{AM,GV,Landau1,Landau2,Schofield}. One of the beauties of Fermi liquid theory is that it reduces the description of the interacting electron fluid to a small number of (Landau) parameters. Therefore, ratios in which these parameters cancel, such as the Wilson-Sommerfeld ratio and Wiedemann-Franz law  \cite{AM,Hewson} provided important tests of Fermi liquid theory. 

In a Fermi liquid  the electronic  contributions to the resistivity [$\rho_{el}\left(T\right)=AT^2$] and heat capacity [$C_{el}\left(T\right)=\gamma T$] are both governed by the effective mass, $m^*$ -- roughly speaking $A\propto m^{*2}$ and $\gamma\propto m^*$. So the Kadowaki-Woods ratio, $A/\gamma^2$, should be  constant in a Fermi liquid  \cite{AK,PAH,Coleman,Auerbach,Miyake}.
More precisely one might expect correlations to leave the Kadowaki-Woods ratio (KWR) unrenormalized because a Kramers-Kronig transformation  relates the real and imaginary parts of the self-energy  \cite{Luttinger,Miyake,Jacko}, which determine the electronic contributions to the heat capacity and resistivity respectively. This means that the KWR is somewhat similar to a fluctuation-dissipation theorem.
 
First Rice  \cite{Rice} and later Kadowaki and Woods  \cite{KW} found that $A/\gamma^2$ is approximately constant within classes of  materials (transition metals and heavy fermion compounds, respectively). However, the ratio differs by two orders of magnitude between these two classes. 
It was subsequently discovered that the Kadowaki-Woods ratio (KWR)  in transition metals and organic charge transfer salts can be even larger than in the heavy fermions  (see Refs. \onlinecite{Jacko,Hussey} and references therein.)

It was long believed  \cite{Miyake,Li} that the size of the KWR gave an indication of the strength of the electron-electron scattering \footnote{Miyake \etal  \cite{Miyake} argued that variations in the KWR arise from differences in the quasiparticle weight.}. 
However, this has been shown to be incorrect \cite{Jacko,Hussey}. Rather, the large variations in the KWR between different classes of materials can be explained almost entirely by taking into account non-interacting properties of the materials (e.g. electron density and dimensionality) \cite{Jacko,Hussey}. Furthermore, it has been shown \cite{Jacko} that the modified KWR $A/\gamma^2f$, where $f$ is a material specific function of the non-interacting band structure (defined below), takes the same predicted value, $81/4\pi\hbar k_B^2e^2$, in a large range of transition metals, charge transfer salts, heavy fermion compounds, and elemental metals, a result which has since been  verified in many other materials  \cite{Jref1,Jref2,Jref3,Jref4,Jref5}. 

Deviations from this universal predicted value of the KWR could provide an indication of non-Fermi liquid behavior. 
Most previous calculations  of the (modified) KWR \cite{Jacko,PAH,Auerbach} have focused on simple, single-band models with toy dispersion relations, e.g. spherical Fermi surfaces. 
Understanding the effects of more complicated and realistic band structures is important if the modified KWR is to be used to identify deviations from Fermi liquid theory.

Previous studies of the KWR in systems with orbital degeneracy \cite{Kontani,Tsuji} or multiple bands \cite{Hussey} have found that either of these can cause significant variation in the KWR. In this paper, we derive a modifed KWR for Fermi liquids with arbitrary dispersion relations, including band structures with multiple bands. We find that the universality of the KWR evident in the single band expression is not a general feature of the multiple band case. In particular in the most general case the strength of electronic correlations {\it does} affect the value of the KWR. This is extremely puzzling as the KWR is found to be close to its universal value in many multiband systems. However, if the renormalization is the same on all bands (in a sense made precise below) correlations do cancel from the KWR.

 In the case of uniform renormalization across all bands, we  demonstrate, that the KWR $R_{KW}\propto1/nN_b^2$ for massive quasiparticles in a system of $N_b$ bands. This is particularly relevant to  semimetals where the low career density opens the possibility of large variations in the career density, $n$. This is further enhanced in   Dirac semimetals -- we find that the massless fermion dispersion relations lead to $R_{KW}\propto1/n^2N_b^2$ in three dimensions. In two dimensions a similar result holds once the KWR is generalized  to account for the logarithmic factor in the imaginary part of the self-energy. Furthermore, we  show that the  logarithmic factor in the imaginary part of the self-energy leads to an increase in the KWR, which may provide an experimental signature of this factor in the self energy.

The remainder of this paper is laid out as follows.
In the following section, we calculate the resistivity of an arbitrary multi-band Fermi liquid by calculating the conductivity from the Kubo formula of linear response. In Section \ref{heat}, we calculate the effect of multiple bands on the heat capacity. In Section \ref{KWRSec}, we combine these results to determine the form of the Kadowaki-Woods ratio in arbitrary band structuress. In Section \ref{graphenesec}, we then apply this form for the KWR to simple models of Dirac semimetals in two and three dimensions.

\section{Conductivity from the Kubo Formula}

In general, the contribution to the intraband self-energy from interband terms scales quadratically with the intraband self-energy, such that, as long as the self-energy is small, the intraband contribution will dominate
\footnote{For a derivation see the supplementary material.}. When intraband scattering is the dominant contribution to the scattering rate, the diagonal component of the conductivity tensor, $\sigma_{xx}$, for a material with $N_b$ bands crossing the Fermi surface in the low temperature limit is \footnote{The derivation of Eq. (\ref{eqn:dc}) is a straightforward generalization of the textbook one band case, but we could not find it in the literature and therefore we give the derivation in the supplementary information.}
\begin{eqnarray}
\sigma_{xx}&=&e^2\hbar\int\limits_{-\infty}^{\infty}\frac{d^3\textbf{k}}{\left(2\pi\right)^3} \int\limits_{-\infty}^{\infty} \frac{d\omega}{4\pi} \sum\limits_{\tilde{a}}^{2N_b} \frac{2\pi Z_{\tilde{a}}\delta\left(\omega-Z_{\tilde{a}} \omega_{\textbf{k}\tilde{a}}\right)} {\Sigma_{\tilde{a}\tilde{a}}''\left(\omega\right)}v_{\textbf{k}x\tilde{a}}^2 \left(\frac{dn_f(\omega)}{d\omega}\right), \label{eqn:dc}
\end{eqnarray}
where $\Sigma_{\tilde{a}\tilde{a}}\left(\omega\right)=\Sigma_{\tilde{a}\tilde{a}}'\left(\omega\right)+i\Sigma_{\tilde{a}\tilde{a}}''\left(\omega\right) $ is the  self-energy, $Z_{\tilde{a}}=|1-[\partial\Sigma'_{\tilde{a}\tilde{a}}\left(\omega\right)/\partial\omega]_{\omega=0}|^{-1}$ is the quasiparticle weight,  $v_{\textbf{k}x\tilde{a}} $ the $x$-component of the group velocity of an electron in spin-band $\tilde{a}=\left(a,\sigma\right)$, where $a$ denotes the band and $\sigma$ the spin (giving $2N_b$ spin-bands), and we have neglected vertex corrections (cf. Ref. \onlinecite{Jacko}). The sharply peaked derivative of the Fermi-Dirac distribution at low temperatures implies that
\begin{eqnarray}
\sigma_{xx}&\approx &\sum\limits_{\tilde{a}=1}^{2N_b}\frac{e^2\hbar\langle v_{\textbf{k}x\tilde{a}}^2\rangle}{2} \int\limits_{-\infty}^{\infty}\frac{d^3\textbf{k}}{\left(2\pi\right)^3} Z_{\tilde{a}}\delta\left(Z_{\tilde{a}}\mu-Z_{\tilde{a}} \omega_{\textbf{k}\tilde{a}}\right) \int\limits_{-\infty}^{\infty} d\omega \left(\frac{-1}{\Sigma_{\tilde{a}\tilde{a}}''\left(\omega\right)} \right)\left(-\frac{dn_f(\omega)}{d\omega}\right),\label{2bc}
\end{eqnarray}
where $\langle ... \rangle$ indicates an average over the Fermi surface. The conductivity is clearly then the sum, in series, of the conductivities of the individual bands:
\begin{eqnarray}
\sigma_{xx}=\sum\limits_{\tilde{a}=1}^{2N_b} \sigma_{xx\tilde{a}}&=&\sum\limits_{\tilde{a}=1}^{2N_b}\frac{e^2\hbar\langle v_{\textbf{k}x\tilde{a}}^2\rangle}{2} D_{0;\tilde{a}}  \int\limits_{-\infty}^{\infty} d\omega   
\left(\frac{1}{\Sigma_{\tilde{a}\tilde{a}}''\left(\omega\right)}\right) \left(\frac{dn_f(\omega)}{d\omega}\right),\label{2bcf}
\end{eqnarray}
where the bare density of states of band $\tilde{a}$ at the Fermi level is $D_{0;\tilde{a}}= (2\pi)^{-3}\int_{-\infty}^{\infty}d^3\textbf{k} \left[\delta\left(\omega_{\textbf{k}\tilde{a}}-\mu\right)\right]$ and $\sigma_{xx\tilde{a}}$ is the conductivity of due to spin-band $\tilde{a}$. It immediately follows that, if interband scattering is neglected, the resistivities of the bands must add in parallel,  consistent with Matthiessen's rule \cite{AM}.

\subsubsection{The Quadratic Contribution to the Resistivity}

To determine the explicit form of the conductivity in a Fermi liquid, we generalise a local (momentum-independent) phenomenological self-energy model proposed by Miyake, Matsuura and Varma, \cite{Miyake} to include multiple Fermi surfaces. We assume that in each band the self-energy takes the phenomenological form
\begin{subequations}
\begin{eqnarray}
\Sigma''_{\tilde{a}\tilde{a}}\left(\omega\right)&=& -\frac{\hbar}{2\tau_{0;\tilde{a}}} -s_{\tilde{a}}\frac{\omega^2+\left(\pi k_BT\right)^2}{\left(\omega^*_{\tilde{a}}\right)^2} \hspace*{2cm} 
\text{ for } \left|\omega^2+\left(\pi k_BT\right)^2\right|<\left(\omega^*_{\tilde{a}}\right)^2 \text{ and } \\
&=& -\left(\frac{\hbar}{2\tau_{0;\tilde{a}}} +s_{\tilde{a}}\right)F\left[\left(\frac{\omega^2+\left(\pi k_BT\right)^2}{\left(\omega^*_{\tilde{a}}\right)^2}\right)^{\frac{1}{2}}\right] 
\text{ for } \left|\omega^2+\left(\pi k_BT\right)^2\right|>\left(\omega^*_{\tilde{a}}\right)^2, 
\end{eqnarray}\label{phenom}\end{subequations}
where the energy $\omega$ is measured from the Fermi level, $\tau_{0;\tilde{a}}^{-1}$ is the impurity scattering rate, $s_{\tilde{a}}=n_{\tilde{a}}/3\pi D_{0;\tilde{a}}$ is the electron-electron scattering rate in the unitary limit \cite{Miyake,Jacko} (with $n_{\tilde{a}}$ the  density of charge carriers in band $\tilde{a}$), $F(x)$ is a monotonically decreasing function with $F(1)=1$ and $F(\infty)=0$ and $\omega^*_{\tilde{a}}$ is an energy scale characterising the strength of the many-body correlations (we will show below that $Z_{\tilde{a}} \approx \omega^*_{\tilde{a}} / 4s_{\tilde{a}}$). This is the natural form for a local Fermi liquid provided interband interactions are weak compared to the intraband interactions. If the two are comparable (as in some multiorbital models) the self energy is proportional to $N_b-1$ \cite{Kontani}.

In the limit of vanishing impurity scattering, $\tau^{-1}_{0;\tilde{a}}\rightarrow 0$, the conductivity is then given by
\begin{eqnarray}
\sigma_{xx}&=&-\sum\limits_{\tilde{a}=1}^{2N_b}\frac{3\pi e^2\hbar }{2}\frac{D_{0;\tilde{a}}^2\langle v_{0x\tilde{a}}^2\rangle\left(\omega^*_{\tilde{a}}\right)^2}{n_{\tilde{a}}} 
\int\limits_{-\infty}^{\infty} d\omega   
\frac{1}{\omega^2+\left(\pi k_BT\right)^2}\frac{dn_f(\omega)}{d\omega}.\label{2bcondlowT}
\end{eqnarray}
After computing the energy integral, 
the $N_b$-band $A$ coefficient is
\begin{eqnarray}
A&=&\frac{8 k_B^2}{\pi e^2\hbar}\left(\sum\limits_{\tilde{a}=1}^{2N_b}\langle v_{0x\tilde{a}}^2\rangle\frac{D_{0;\tilde{a}}^2\left(\omega^*_{\tilde{a}}\right)^2}{n_{\tilde{a}}} \right)^{-1}
=\left(\sum\limits_{\tilde{a}=1}^{2N_b} A_{\tilde{a}}^{-1}\right)^{-1},
\label{Ainparallel}
\end{eqnarray}
thus, we see that the coefficients of the individual bands, $A_{\tilde{a}}$, add in parallel.

\section{The Heat Capacity via Kramers-Kronig transform}\label{heat}

It follows from the extensivity of the heat capacity that, in a multi-band system, the total heat capacity is given by the sum in series of the heat capacity due to each individual band \cite{AM},
\begin{eqnarray}
\gamma&=&\frac{C_{el; N_b}\left(T\right)}{T} =\frac{\pi^2 k_B^2}{3} \sum\limits_{\tilde{a}=1}^{2N_b} \frac{D_{0;\tilde{a}}}{Z_{\tilde{a}}}
=\frac{\pi^2 k_B^2}{3}\sum\limits_{\tilde{a}=1}^{2N_b} D_{0;\tilde{a}}\left(1-\frac{\partial \Sigma'_{\tilde{a}\tilde{a}}\left(\omega\right)}{\partial \omega}\right). \label{2bC}
\end{eqnarray}

To determine how the heat capacity is influenced by interactions, and calculate the relevant $\gamma$ coefficients, we need to first find the real part of the self-energy in each band. The Fermi liquid self-energy, being a causal response function in the time domain, satisfies the conditions for the Kramers-Kronig relations in the frequency domain  \cite{Luttinger} -- in particular $\lim\limits_{\omega\rightarrow\infty} \Sigma''\left(\omega\right)\rightarrow 0$. Knowledge of the form of the imaginary part of the self-energy is therefore sufficient to determine the real part, which appears in the definition of the quasiparticle weight and therefore in the expression for the heat capacity. The real part of the self-energy within each band is then
\begin{eqnarray}
\Sigma'_{\tilde{a}\tilde{a}}\left(\omega\right)&=&\frac{1}{\pi} P \int\limits_{-\infty}^{\infty} d\omega' \frac{\Sigma''_{\tilde{a}\tilde{a}}(\omega')}{\omega'-\omega}\nonumber\\
&=& \frac{-s_{\tilde{a}}}{\pi} \left[\vphantom{P \int\limits_{-\omega^*_{\tilde{a}}}^{\omega^*_{\tilde{a}}} \frac{d\omega'}{\omega'-\omega} \left(\frac{\omega'^2}{\left(\omega^*_{\tilde{a}}\right)^2}\right)}
P\int\limits_{-\infty}^{-\omega^*_{\tilde{a}}} \frac{d\omega'}{\omega'-\omega} F\left[\frac{\left|\omega'\right|}{\omega^*_{\tilde{a}}}\right]
+P \int\limits_{-\omega^*_{\tilde{a}}}^{\omega^*_{\tilde{a}}} \frac{d\omega'}{\omega'-\omega} \left(\frac{\omega'^2}{\left(\omega^*_{\tilde{a}}\right)^2}\right) 
 +P\int\limits_{\omega^*_{\tilde{a}}}^{\infty} \frac{d\omega'}{\omega'-\omega} F\left[\frac{\left|\omega'\right|}{\omega^*_{\tilde{a}}}\right]\right] ,
\label{KKint}
\end{eqnarray}
where we have again taken the limit of vanishing impurity scattering, and have restricted the pole of the integral, $\omega'=\omega$, to occur below the cut-off energy scale, $|\omega'|\le|\omega^*_{\tilde{a}}|$ (i.e., only considered low-energy excitations). The first term in Eq. (\ref{KKint}) contributes a logarithmic term to the result, which we approximate by the lowest order terms in a Taylor series expansion, while the second and third terms contribute linearly to the self-energy. Neglecting terms of order ${\cal O}\left(\omega/\omega^*_{\tilde{a}}\right)^3$ and higher, we find that the real part of the self-energy for a low-energy quasiparticle in band $\tilde{a}$ is
\begin{equation}
\Sigma'_{\tilde{a}\tilde{a}}\left(\omega\right)=-\frac{4n_{\tilde{a}}\omega} {3\pi^2 D_{0;\tilde{a}}\omega^*_{\tilde{a}}} \xi, \label{realse}
\end{equation}
where 
\begin{eqnarray}
\xi&=&\frac{1}{2}\left(1+\int\limits_{1}^{\infty}dy  \frac{F(y)}{y^{2}}\right) \label{xidef}
\end{eqnarray}
it follows straightforwardly from the definition of $F(y)$ that  $1/2\leq\xi\leq 1$, henceforth we take $\xi=1$ for simplicity. Inserting this expression for the real parts of the self-energies into the heat capacity expression Eq. (\ref{2bC}), and taking the strong scattering ($4s_{p\tilde{a}}/\pi\omega^*_{\tilde{a}} \gg1$) limit, which corresponds physically to  $m^*\gg m_0$, we obtain
\begin{eqnarray}
\gamma&=&\frac{4k_B^2}{9} \sum\limits_{\tilde{a}=1}^{2N_b} \frac{n_{\tilde{a}}}{\omega^*_{\tilde{a}}}. \label{2by}
\end{eqnarray}
 Note that  Eqs. (\ref{2bC}) and (\ref{2by}) imply that $\omega^*_{\tilde{a}}\approx4s_{\tilde{a}}Z_{\tilde{a}}=4Z_{\tilde{a}}n_{\tilde{a}}/3\pi D_{0;\tilde{a}}$, which gives a straightforward interpretation of this energy scale.

\section{The Kadowaki-Woods Ratio}\label{KWRSec}

Given the above calculations of $A$ and $\gamma$ for the multiple band system, Eqs. (\ref{Ainparallel}) and (\ref{2by}), we find that the KWR is given by
\begin{widetext}
\begin{eqnarray}
R_{KW;N_b}=\frac{A}{\gamma^2} &=&\frac{81}{2\pi e^2\hbar k_B^2 }\left(\sum\limits_{\tilde{b}=1}^{2N_b}\langle v_{0x\tilde{b}}^2\rangle  \frac{D_{0;\tilde{b}}^2\left(\omega^*_{\tilde{b}}\right)^2}{n_{\tilde{b}}} \right)^{-1}\left[\sum\limits_{\tilde{a}=1}^{2N_b}\frac{n_{\tilde{a}}}{\omega^*_{\tilde{a}}}\right]^{-2}.\label{NbKWR}
\end{eqnarray}
Alternatively, one may write
\begin{eqnarray}
\frac{A}{\gamma^2} f_{x;N_b}\left(\left\lbrace\omega^*_{\tilde{a}}\right\rbrace, \left\lbrace n_{\tilde{a}}\right\rbrace\right)&=&\frac{81}{4\pi e^2 \hbar k_B^2} ,\label{KWRNband}
\end{eqnarray}
where we have defined the material specific function for an $N_b$-band Fermi liquid with the resistivity measured in the $x$ direction,
\begin{eqnarray}
f_{x;N_b}\left(\left\lbrace\omega^*_{\tilde{a}}\right\rbrace, \left\lbrace n_{\tilde{a}}\right\rbrace\right)&=& \frac{1}{2}\left[\sum\limits_{\tilde{a}=1}^{2N_b} \frac{n_{\tilde{a}}}{\omega^*_{\tilde{a}}}\right]^{2} \left[\sum\limits_{\tilde{b}=1}^{2N_b}\langle v_{0x\tilde{b}}^2\rangle  \frac{D_{0;\tilde{b}}^2\left(\omega^*_{\tilde{b}}\right)^2}{n_{\tilde{b}}} \right].\label{Nbfdx}
\end{eqnarray}
\end{widetext}

In the one-band limit, this expression simplifies to that calculated in Ref. \onlinecite{Jacko} (with $n_\uparrow=n_\downarrow=n/2$ and $D_{0,\uparrow}=D_{0,\downarrow}=D_0/2$)
\begin{eqnarray}
f_{x;1}\left(n\right)&=&  n \langle v_{0x}^2\rangle D_0^2 \label{1bfdxcomp}.
\end{eqnarray}

The inclusion of multiple Fermi surface sheets significantly complicates the form of $f_{x;N_b}\left(\left\lbrace\omega^*_{\tilde{a}}\right\rbrace, \left\lbrace n_{\tilde{a}}\right\rbrace\right)$. Most importantly  $\omega^*_{\tilde{a}}$, which describes the electronic correlations, does not cancel out of the multiband expression as it does for the single band KWR \cite{Jacko}. Therefore, our calculation predicts that the Kadowaki-Woods ratio is not, in general, independent of electronic correlations. This is rather surprising as observed values of the KWR (including the values for many multiband systems)   are in almost universal agreement with the prediction from the single band calculation, that electronic correlations do not influence the KWR \cite{Jacko}. 

It is therefore important to ask how renormalization effects might cancel in the multiband case and hence universality might be recovered.


The simplest limiting case for which  the effects of many-body correlations cancel out of the KWR is when $\omega_{\tilde{a}}^*$ is independent of the band index, $\tilde{a}$. This is a straightforward extension of the earlier assumption of the locality of the self-energy, by assuming that it is independent of band index as well as momentum. This assumption yields
\begin{eqnarray}
f_{x;N_b}\left(\left\lbrace n_{\tilde{a}}\right\rbrace\right)&=& \frac{1}{2}\left[\sum\limits_{\tilde{a}=1}^{2N_b} n_{\tilde{a}}\right]^{2} \left[\sum\limits_{\tilde{b}=1}^{2N_b}\langle v_{0x\tilde{b}}^2\rangle  \frac{D_{0;\tilde{b}}^2}{n_{\tilde{b}}} \right]. \label{Nbfsimple}
\end{eqnarray} 
Though this calculation has been performed with exactly uniform correlation strengths for simplicity, the result will hold approximately while the correlation strengths are close to uniform. 

Other limits do produce a universal KWR, for example, for free fermions in three dimensions $D_{0;\tilde{b}}\propto n_{\tilde{b}}$, then if the carrier density in one band is much larger than all others the correlations cancel from the KWR. But it seems unlikely that this is relevant the behaviour of a broad range range of materials.
Note, in particular, that if we have a single heavy band it will dominate the heat capacity, but be shorted out of the resistivity. Therefore, the limit of a single heavy band is far from universal.
Therefore, the above calculation seems to suggest the correlation strength (as measured by $\omega_{\tilde{a}}^*=4Z_{\tilde{a}}n_{\tilde{a}}/3\pi D_{0;\tilde{a}}$) does not vary strongly between different bands in strongly correlated systems.


If, further, all of the bands are identical, i.e., if the carrier density $n$, the  Fermi velocity  $\langle v_{0x}^2\rangle$, and density of states $D_0$ are equal for all bands, we have 
\begin{eqnarray}
f_{x;N_b}\left(n\right)&=& N_b^2\langle v_{0x}^2\rangle n D_0^2=N_b^2f_{dx;1}(n),\label{funif}
\end{eqnarray}
and
\begin{eqnarray}
R_{KW;N_b}&=&\frac{81}{4N_b^2\pi e^2\hbar k_B^2\langle v_{0x}^2\rangle n D_0^2}=\frac{R_{KW;1}}{N_b^2}.
\label{2bKWRnDeq} 
\end{eqnarray}
Thus we see that the expression for the KWR in the single-band case is modified by a simple factor of $1/N_b^2$. 

At first glance this expression appears rather similar to the finding of Kontani \etal \cite{Kontani,Tsuji} that in the 
multiorbital periodic Anderson model with $N_o$ impurity states the KWR is reduced by a factor of $1/N_o(N_o-1)$. However, closer examination reveals that the results are actually very different. 
In particular Kontani's factor arises because of interorbital terms in the self energy whereas our $1/N_b^2$ factor arises purely from the electronic structure. We do not obtain Kontani's factor because of our assumption that interband interactions are irrelevant at low energies. In contrast the model Hamiltonian studied by Kontani \cite{Kontani} explicitly sets the intra- and inter-orbital interactions to the same strength. Which approach is appropriate will depend on the material. This therefore adds another layer of non-universality to the KWR.

\section{Dirac Semimetals}
\label{graphenesec}

The general expression, Eq. (\ref{KWRNband}), can be applied to systems of arbitrary band structure to calculate the generalised Kadowaki-Woods ratio, taking into account the effects of multiple bands. Even for materials where the quasiparticle weight is similar for all bands such efforts will, in general, involve first principles band structure calculations. 
In this section, we calculate the KWR for  a simple, linear dispersion ($\varepsilon_{\textbf{k}\tilde{a}}=\hbar v_F \left|\textbf{k}\right|$) appropriate for Dirac semimetals. These models present analytically tractable and instructive examples of complicated band structures for which the presence of multiple bands is important (here, the number of bands may be taken as equivalent to the number of Dirac cones, as each cone will form a sheet in the Fermi surface). Furthermore, these materials are an example of materials where the different sheets of the Fermi surface where we expect the correlations to affect all bands equally and the low carrier density suggests that correlations will play an important role. We also determine an expression for the KWR in two dimensions, accounting for the logarithmic factor arising in the self-energy  \cite{FLdopedgraphene,GrRMP1,GrRMP2,GrRMP3,gbandrenorm,gproplong,Chaplik,Wilkins,SE2D,UC}. 

\subsection{3D Dirac Semimetals} 

We first apply the KWR expression to a simple model of a two-band three-dimensional Dirac semimetal, e.g. Cd$_3$As$_2$, which possesses properties similar to doped graphene \cite{Cd3As2-1,Cd3As2-2}. 
The spatial symmetry between the two bands and the spin symmetry
simplifies the calculation greatly; applying the expression for $N_b$ identical bands, Eq. (\ref{funif}), we find 
\begin{equation}
\frac{A}{\gamma^2} = \frac{1}{\tilde{f}_{3x,2}}\frac{81}{4\pi e^2 \hbar k_B^2}  = \frac{81\hbar}{8e^2k_B^2 n^2}, \label{grapheneKWR} 
\end{equation}
where $\tilde{f}_{3x,2} = 2 n^2/\pi \hbar^2$ is the material specific function for a 3D Dirac semimetal and is larger by a factor of 4 than the corresponding $f$ in a single-band calculation with the same dispersion. We note that the Kadowaki-Woods ratio for these materials depends straightforwardly on the electronic density, which is tunable via chemical doping  \cite{Cd3As2-1,Cd3As2-2}, providing a potential experimental test of this expression.

\subsection{Doped Graphene and the Self-Energy in Two Dimensions}

In calculating Eq. (\ref{grapheneKWR}), we utilised the form of self-energy given in Eq. (\ref{phenom}), which implicitly assumes a three-dimensional material. In two dimensions, the self-energy differs from the three dimensional case, with the inclusion of an additional logarithmic factor \cite{2DEG1,2DEG2,2DEG3,dasSarmaSE,dasSarmaSE2,FLTin2D,GQ,GV}. In order to calculate the KWR for doped graphene, we introduce the following model for the imaginary part of the self-energy at low energies and temperatures for 2D systems
\begin{eqnarray}
\Sigma''_{\tilde{a}\tilde{a}}\left(\omega\right)&=& s_{\tilde{a}}\frac{\omega^2+\left(\pi k_B T\right)^2}{\left(\omega^*_{\tilde{a}}\right)^2}\log\left(\frac{\sqrt{\omega^2+\left(\pi k_B T\right)^2}}{B_{\tilde{a}}\omega^*_{\tilde{a}}}\right),
\label{gSim}
\end{eqnarray}
where $B_{\tilde{a}}$ is a constant of order unity (for example the calculation in Ref. \onlinecite{dasSarmaSE2} gives $B_{\tilde{a}}\approx1/\pi$ for graphene; but other calculations give slightly different results). We stress that, in the zero and high temperature ($k_BT\gg\omega$) limits, Eq. (\ref{gSim}) reproduces the  known results for those limits \cite{2DEG1,dasSarmaSE,dasSarmaSE2}. We further assume that, above the relevant energy scale, $\omega^*$, the self-energy decreases monotonically, as in the 3D case. From this expression for imaginary part of the self-energy, we calculate the real part using the Kramers-Kronig transformation, and find that
\begin{eqnarray}
\Sigma'_{\tilde{a}\tilde{a}}\left(\omega\right)&=&-\frac{4s}{\pi}\left[\xi\log\left(B_{\tilde{a}}\right)+1\right]\omega+ {\cal O}\left(\omega^3\right)\label{gSre}
\end{eqnarray}
where the logarithmic factor arises due to the requirement that the self-energy be continuous at $\omega=\omega^*$. 


In defining the 2D KWR, we must account for the fact that the logarithmic contribution to the imaginary part of the self-energy (and scattering rate) results in a corresponding logarithmic factor in the resistivity, $\rho=-\tilde{A}T^2\log\left(\pi k_BT/B\omega^*\right)$. The coefficient of this logarithmically adjusted quadratic term will be used in our expression. Calculating the conductivity from the Kubo formula gives
\begin{eqnarray}
\tilde{A}_{\tilde{a}}&=&  \frac{8 k_B^2}{\pi e^2\hbar n_{\tilde{a}}\langle v_{x\tilde{a}}^2\rangle\omega^*_{\tilde{a}} D_{0;\tilde{a}}^2}\label{2DA},
\end{eqnarray}
where, in evaluating the energy integral (see Eq. (\ref{2bcondlowT})), we have approximated the logarithmic contribution by its value at the Fermi surface. From the derivative of the real part of the self-energy, we find the value for the quasiparticle weight and therefore the linear coefficient of the heat capacity:
\begin{eqnarray}
\gamma_{\tilde{a}} &=& \frac{2\pi k_B^2n_{\tilde{a}}}{9\pi} \left( \log\left(B_{\tilde{a}}\right) +\frac{1}{2}\right).
\end{eqnarray}

Defining the two-dimensional Kadowaki-Woods ratio as $\tilde{A}/\gamma^2$ we find 
\begin{eqnarray}
R_{KW}=\frac{\tilde{A}}{\gamma^2}&=&\frac{81}{4\pi e^2\hbar k_B^2}  \frac{1}{\tilde{f}_{2x;2}\left(n\right)}
\end{eqnarray}
where
\begin{widetext}
\begin{eqnarray}
\tilde{f}_{2x;2}\left(n\right)&=& \frac{1}{2}\left[\sum\limits_{\tilde{a}=1}^{2N_b} \frac{n_{\tilde{a}}\left( \log\left(B_{\tilde{a}}\right) +\frac{1}{2}\right)}{\omega^*_{\tilde{a}}}\right]^{2} \left[\sum\limits_{\tilde{b}=1}^{2N_b}  \frac{\langle v_{0x\tilde{b}}^2\rangle D_{0;\tilde{b}}^2\left(\omega^*_{\tilde{b}}\right)^2}{n_{\tilde{b}}C_{\tilde{b}}} \right]\nonumber
\end{eqnarray}
\end{widetext}
is the two-dimensional material specific function, which for graphene takes the form
\begin{eqnarray}
\tilde{f}_{2x;2}\left(n\right)=4\tilde{f}_{2x;1}\left(n\right)&=&\frac{2n^2}{\pi\hbar^2}\left[\log\left(B\right)+\frac{1}{2}\right]^2.
\end{eqnarray}
Here the Kadowaki-Woods ratio again depends on the electron density in a straightforward manner, but differs from the previously derived expression (Eq. (\ref{grapheneKWR})) only by a factor of $\left[ \log\left(B\right)+1/2\right]^2$. 
Assuming $B \simeq 1/\pi$, this term more than doubles the expected Kadowaki-Woods ratio. Measuring the KWR of graphene presents a number of technical challenges, chiefly the difficulty of performing calorimetric measurements on a material of single atom thickness. It has been a long standing problem to observe the logarithmic factor in the resistivity of a two-dimensional Fermi liquid because it would require data over many orders of magnitude in temperature. However, observation of this correction offers a potential test of the dimensionality of the Fermi liquid self energy without the need to take data over multiple decades.

\section{Conclusions}

We have shown that, in general, the Kadowaki-Woods ratio of a multiband local Fermi liquid is changed by electronic correlations.  
This is in marked contrast to the single band case, where the  KWR is independent of the strength of the electronic correlations. 
It is therefore puzzling that the experimental data suggest the within classes of materials the KWR is 
remarkably consistent, and that the modified KWR is remarkably consistent across many chemically diverse strongly correlated metals. The simplest explanation is that the correlations are indeed very similar across all bands in these materials.
We have also shown that a non-parabolic dispersion does not significantly alter the form of the KWR, provided the fermions remain massive.

In the case of uniform renormalization across bands, we have further demonstrated that  $R_{KW}\propto1/nN_b^2$ for massive quasiparticles in a system of $N_b$ bands. This is particularly interesting in  semimetals where the low career density opens the possibility of large variations in the career density, $n$. This is further enhanced in   Dirac semimetals, where we have shown that the massless fermion dispersion relations lead to $R_{KW}\propto1/n^2N_b^2$ in three dimensions. In two dimensions a similar result holds once the KWR is generalized  to account for the logarithmic factor in the imaginary part of the self-energy. Furthermore, we have shown that the  logarithmic factor in the imaginary part of the self-energy leads to an increase in the KWR, which may provide an experimental signature of this factor in the self energy.
 
\section*{Acknowledgments}

This work was supported by the Australian Research Council (ARC) under grant DP130100757. BJP is supported by the ARC under grant FT130100161.

\bibliography{KWRRef}
\end{document}


\title{When is the Kadowaki-Woods ratio universal?: Supplementary Material}

\maketitle

\section{Scattering and the Self-Energy in Arbitrary Bandstructures Systems}

We wish to determine the form of the self-energy in arbitrary bandstructures and the effect of multiple Fermi surfaces on the self-energy and Green's function. The elements of the dressed $N_b \times N_b$ Green's function matrix are then defined by the matrix form of Dyson's equation \cite{Mahan} and we have (here and below, we use the convention $b\ne a$ to distinguish intraband and interband contributions)
\begin{eqnarray}
G_{a,a}^{-1}\left(\omega\right)&=&\left(\left[G_{a,a}^{(0)}\left(\omega\right)\right]^{-1}- \Sigma_{a,a}\left(\omega\right)\right)_{a,a}, \nonumber \\
G_{a,b}^{-1}\left(\omega\right)&=&- \Sigma_{a,b}\left(\omega\right). \nonumber
\end{eqnarray}
%
The inverse can be calculated using the adjugate matrix \cite{Arfken,Boas} ($\textbf{A}^{-1}=\frac{1}{\left|\textbf{A}\right|}\left(\text{adj}\left[\textbf{A}\right]\right)$ ), each element of which contains $N_b-1$ terms, with a leading term of the form $\prod\limits_{a=1}^{N_b} G_{a,a}^{-1}\left(\omega\right)$, with all other terms of order ${\cal O} \left(\Sigma_{a,b}^2\left(\omega\right) \right)$ and higher. So, with minimal interband scattering, the intraband Green's function elements will be dominated by the intraband self-energy contributions, and the individual bands will reduce to separable channels. We consider in detail the two-band case below, to give an indication of how this separability arises.\footnote{It should be noted that, for systems of $N_b>2$, extra terms arising in the Green's function elements (Eqns. (\ref{G11}) and (\ref{G12})) will be of equal or higher order in the interband self-energy than those in the two-band case, and will contribute progressively less.}
\subsection{The Two-Band Case}
The inverse of the two-band Green's function matrix has the simple representation
\begin{eqnarray}
G\left(\omega\right)&=&\frac{1}{\left|\left[G\left(\omega\right)\right]^{-1}\right|} \begin{bmatrix}
G_2^{-1}\left(\omega\right) & \Sigma_{1,2}\left(\omega\right) \\
\Sigma_{2,1}\left(\omega\right) & G_1^{-1}\left(\omega\right)
\end{bmatrix}, \label{g-1}
\end{eqnarray}
where $G_a^{-1}\left(\omega\right) =\left[G_{a,a}^{(0)}\left(\omega\right)\right]^{-1}-\Sigma_{a,a}\left(\omega\right)$ is the dressed Green's function in band $a$ neglecting the effects of interband scattering (i.e. $\Sigma_{a,b}\left(\omega\right)=0$), and the determinant is
\begin{eqnarray}
\left|\left[G\left(\omega\right)\right]^{-1}\right|&=&\prod\limits_{a=1}^{2}\left[G_{a,a}\left(\omega\right)\right]^{-1}-\Sigma_{1,2}\left(\omega\right)\Sigma_{2,1}\left(\omega\right).
\label{det}
\end{eqnarray}

The intraband elements of the dressed Green's function matrix are then given by 
\begin{eqnarray}
G_{a,a}\left(\omega\right)&=& \left[ G_{a}^{-1}\left(\omega\right)-\frac{\Sigma_{a,b}\left(\omega\right)\Sigma_{b,a}\left(\omega\right)} {G_{b}^{-1}\left(\omega\right)}\right]^{-1},\label{G11}
\end{eqnarray}
and the interband elements by
\begin{eqnarray}
G_{a,b}\left(\omega\right)&=&\left[\frac{\prod\limits_{n=1}^{2}G_{n}^{-1}\left(\omega\right)} {\Sigma_{a,b}\left(\omega\right)} -\Sigma_{b,a}\left(\omega\right)\right]^{-1}. 
\label{G12}
\end{eqnarray}

The interband self-energy elements appear here as corrections to the single-band form of the intraband Green's function elements, $G_{a,a}\left(\omega\right)$. In the limit of small interband scattering, these corrections become negligible, being of order ${\cal O} \left(\left|\Sigma_{1,2}\left(\omega\right)\right|^2\right)$. As the interband elements of the self-energy approach zero, so do the $G_{a,b}\left(\omega\right)$ elements, as interband propagation occurs only in the presence of interband scattering. Therefore, in the limit of vanishing interband scattering, the dressed Green's function matrix is diagonal (i.e. intraband only), 
\begin{eqnarray}
\lim\limits_{\Sigma_{i,j\ne i}\rightarrow 0 } G_{a,a}\left(\omega\right) = G_{a}\left(\omega\right), \nonumber\\
\lim\limits_{\Sigma_{i,j\ne i}\rightarrow 0} G_{a,b}\left(\omega\right) = 0. \nonumber
\end{eqnarray}

The matrix elements Eqns. (\ref{G11}) and (\ref{G12}) can be expanded by expressing each self-energy as a sum of its real and imaginary parts, and making use of the knowledge that the self-energy matrix must be Hermitian (so $\Sigma_{a,b}\left(\omega\right)=\Sigma_{b,a}\left(\omega\right)^\dagger$). The diagonal component is then
\begin{widetext}
\begin{eqnarray}
G_{a,a}\left(\omega\right)&=&\left[ \left[\left[G_{a,a}^{(0)}\left(\omega\right)\right]^{-1}-\Sigma_{a,a}'\left(\omega\right)-i\Sigma_{a,a}''\left(\omega\right) \right] -\frac{\left|\Sigma_{a,b}\left(\omega\right)\right|^2}{\left[\left[G^{(0)}_{b,b}\left(\omega\right)\right]^{-1}-\Sigma'_{b,b}\left(\omega\right)- i\Sigma''_{b,b}\left(\omega\right)\right]}\right]^{-1}.
\end{eqnarray}
\end{widetext}

For notational ease, we define $\eta_{\textbf{k},a}\left(\omega\right)$ (the renormalised real part of the inverse Green's function) and $\Gamma_{a}\left(\omega\right)$ (the inverse quasiparticle lifetime) as 
\begin{eqnarray}
\eta_{\textbf{k},a}\left(\omega \right)&=& \omega-\varepsilon_{\textbf{k},a}- \Sigma'_{a,a}\left(\omega\right)\label{eta2} ,\\
\Gamma_{a}\left(\omega\right) &=& \Sigma''_{a,a}\left(\omega\right)-\delta\label{Gam2},
\end{eqnarray}\\\\
where $\varepsilon_{\textbf{k};a}$ refers to the dispersion for the $a$ band, measured from the Fermi energy. We find a more explicit expression for the intraband Green's function element
\begin{widetext}
\begin{eqnarray}
G_{a,a}\left(\omega\right)&=& \frac{\left[\eta_{\textbf{k},b}^2\left(\omega\right)+ \Gamma_{b}^2\left(\omega\right)\right]\left[\eta_{\textbf{k},a}\left(\omega\right) \left(\eta_{\textbf{k},b}^2\left(\omega\right)+\Gamma_{b}^2\left(\omega\right)\right)- \left|\Sigma_{b,a}\left(\omega\right)\right|^2\eta_{\textbf{k},b}\left(\omega\right)\right]} {\left[\eta_{\textbf{k},a}\left(\omega\right) \left(\eta_{\textbf{k},b}^2\left(\omega\right)+\Gamma_{b}^2\left(\omega\right)\right)- \left|\Sigma_{b,a}\left(\omega\right)\right|^2\eta_{\textbf{k},b}\left(\omega\right)\right]^2+ \left[\Gamma_{a}\left(\omega\right)\left(\eta_{\textbf{k},b}^2\left(\omega\right)+ \Gamma_{b}^2\left(\omega\right)\right)+\left|\Sigma_{b,a}\left(\omega\right)\right|^2 \Gamma_{b}\left(\omega\right)\right]^2}  \nonumber\\
&+& \frac{ i\left[\eta_{\textbf{k},b}^2\left(\omega\right)+\Gamma_{b}^2\left(\omega\right)\right] \left[\Gamma_{a}\left(\omega\right)\left(\eta_{\textbf{k},b}^2\left(\omega\right)+ \Gamma_{b}^2\left(\omega\right)\right)+\left|\Sigma_{b,a}\left(\omega\right)\right|^2 \Gamma_{b}\left(\omega\right)\right]}{\left[\eta_{\textbf{k},a}\left(\omega\right) \left(\eta_{\textbf{k},b}^2\left(\omega\right)+\Gamma_{b}^2\left(\omega\right)\right)- \left|\Sigma_{b,a}\left(\omega\right)\right|^2\eta_{\textbf{k},b}\left(\omega\right)\right]^2+ \left[\Gamma_{a}\left(\omega\right)\left(\eta_{\textbf{k},b}^2\left(\omega\right)+ \Gamma_{b}^2\left(\omega\right)\right)+\left|\Sigma_{b,a}\left(\omega\right)\right|^2 \Gamma_{b}\left(\omega\right)\right]^2},  \nonumber\\ \label{G22new}
\end{eqnarray}
and, for the interband elements of the dressed Green's function matrix,
\begin{eqnarray}
G_{a,b}\left(\omega\right)^{-1}&=& \frac{\prod\limits_{n=1}^{2} \left[\left[G^{(0)}_{n,n}\left(\omega\right)\right]^{-1}- \Sigma'_{n,n}\left(\omega\right)-i\Sigma''_{n,n}\left(\omega\right)\right] -\left|\Sigma_{a,b}\left(\omega\right)\right|^2} {\Sigma'_{a,b}\left(\omega\right)+i\Sigma''_{a,b}\left(\omega\right)}. \nonumber
\end{eqnarray}
Explicitly expanding the product over the bands, we obtain from this the expression for the interband elements of the Green's function
\begin{eqnarray}
G_{a,b}\left(\omega\right)&=&\left[\frac{\Sigma'_{a,b}\left(\omega\right) \left(\eta_{\textbf{k},a}\left(\omega\right)\eta_{\textbf{k},b}\left(\omega\right)- \Gamma_{a}\left(\omega\right)\Gamma_{b}\left(\omega\right) -\left|\Sigma_{a,b}\left(\omega\right)\right|^2\right)-\Sigma''_{a,b}\left(\omega\right) \left(\Gamma_{a}\left(\omega\right)\eta_{\textbf{k},b}\left(\omega\right)+ 
\eta_{\textbf{k},a}\left(\omega\right) \Gamma_{b}\left(\omega\right)\right)}{\left|\Sigma_{a,b}\left(\omega\right)\right|^2}\right. \nonumber\\
&&\left.-\frac{i\left(\left(\Gamma_{a}\left(\omega\right) \eta_{\textbf{k},b}\left(\omega\right)+\eta_{\textbf{k},a}\left(\omega\right) \Gamma_{b}\left(\omega\right)\right)\Sigma'_{a,b}\left(\omega\right)+ \Sigma''_{a,b}\left(\omega\right)\left(\eta_{\textbf{k},a}\left(\omega\right) \eta_{\textbf{k},b}\left(\omega\right)-
 \Gamma_{a}\left(\omega\right)\Gamma_{b}\left(\omega\right) -\left|\Sigma_{a,b}\left(\omega\right)\right|^2\right)\right)} {\left|\Sigma_{a,b}\left(\omega\right)\right|^2}\right]^{-1}, \nonumber\\
\label{G12new}
\end{eqnarray}
\end{widetext}
which, as expected, is proportional to the interband self-energy contribution, so that, in the limit of vanishing interband scattering, the interband Green's function elements vanish also (i.e. $\lim\limits_{\Sigma_{a,b}\rightarrow 0} G_{a,b}=G_{a,b}^{(0)}=0$).
\subsection{The Spectral Density Function}
For the case of a two-band material, we concentrate here on the intraband (diagonal) terms $G_{a,a}\left(\omega\right)$ (Eq. (\ref{G22new})), taking the low energy limit \cite{Mahan}, as the interband terms do not arise in the Kubo formula for conductivity (see below). In the more general case, for small interband scattering similar arguments to those above hold and the individual bands are separable, with the intraband spectral density functions reducing to their individual single-band forms. For the intraband elements in the two-band system, the spectral densities are 
\begin{widetext}
\begin{eqnarray}
A_{\textbf{k},a}\left(\omega\right) &=& \frac{-2\left(\eta_{\textbf{k},b}^2\left(\omega\right)+ 
\Gamma_{b}^2\left(\omega\right)\right) \left[\Gamma_{a}\left(\omega\right)\left(\eta_{\textbf{k},b}^2\left(\omega\right)+ \Gamma_{b}^2\left(\omega\right)\right)+\left|\Sigma_{a,b}\left(\omega\right)\right|^2 \Gamma_{b}\left(\omega\right)\right]}{\left[\eta_{\textbf{k},a}\left(\omega\right) \left(\eta_{\textbf{k},b}^2\left(\omega\right)+\Gamma_{b}^2\left(\omega\right)\right)- \left|\Sigma_{a,b}\left(\omega\right)\right|^2\eta_{\textbf{k},b}\left(\omega\right)\right]^2+\left[\Gamma_{a}\left(\omega\right)\left(\eta_{\textbf{k},b}^2\left(\omega\right)+ \Gamma_{b}^2\left(\omega\right)\right)+ 
\left|\Sigma_{a,b}\left(\omega\right)\right|^2\Gamma_{b}\left(\omega\right)\right]^2}. \nonumber\\
\label{sdf11}
\end{eqnarray}
\end{widetext}
Again, we take the low energy limit, in the two band case given by $\Gamma_{a,b}=\Sigma''_{a,b} \rightarrow 0$, restricting quasiparticles in both bands to be close to the Fermi surface. We discard terms of order ${\cal O} \left(\Sigma_{a,b}''^2\left(\omega\right) \right)$ and higher to obtain 
\begin{widetext}
\begin{eqnarray}
A_{\textbf{k},a}\left(\omega\right) &=& \frac{-2\eta_{\textbf{k},b}^2\left(\omega\right)\left[\Gamma_{a}\left(\omega\right) \eta_{\textbf{k},b}^2\left(\omega\right)+\Sigma_{a,b}'^2\left(\omega\right) \Gamma_{b}\left(\omega\right)\right]}{\left[\eta_{\textbf{k},a}\left(\omega\right)\eta_{\textbf{k},b}^2\left(\omega\right)- \Sigma_{a,b}'^2\left(\omega\right)\eta_{\textbf{k},b}\left(\omega\right)\right]^2+\left[\Gamma_{a}\left(\omega\right)\eta_{\textbf{k},b}^2\left(\omega\right)+ \Sigma'^2_{a,b}\left(\omega\right) \Gamma_{b}\left(\omega\right)\right]^2}.\label{sdf11l}
\end{eqnarray}
\end{widetext}
We can simplify this expression further by assuming that the interband self-energy $\Sigma_{a,b}\left(\omega\right)$ is much smaller than the single band self-energies, which gives the low-energy limits of the spectral density functions,\\
\begin{eqnarray}
\lim\limits_{\Gamma \rightarrow 0} A_{\textbf{k},a}\left(\omega\right) &=& \lim\limits_{\Gamma \rightarrow 0}\frac{-2\eta_{\textbf{k},b}^4\left(\omega\right)\Gamma_{a}\left(\omega\right)} {\left[\eta_{\textbf{k},a}\left(\omega\right)\eta_{\textbf{k},b}^2\left(\omega\right)\right]^2+ \left[\Gamma_{a}\left(\omega\right)\eta_{\textbf{k},b}^2\left(\omega\right)\right]^2} \nonumber\\
&=&\lim\limits_{\Gamma \rightarrow 0}\frac{-2\Gamma_{a}\left(\omega\right)} {\eta_{\textbf{k},a}^2\left(\omega\right)+ \Gamma_{a}^2\left(\omega\right)} \nonumber\\
&=& -2\pi\delta\left(\eta_{\textbf{k},a}\left(\omega\right)\right),\label{sdf11lim}\\
\lim\limits_{\Gamma \rightarrow 0} A^2_{\textbf{k},a}\left(\omega\right) &=& \frac{-2\pi\delta\left(\eta_{\textbf{k},a}\left(\omega\right)\right)} {\Gamma_{a}\left(\omega\right)}. \label{sdf11limsq}
\end{eqnarray}

It can be seen from Eqn. (\ref{sdf11limsq}) that, in the limit of small interband scattering, the spectral density function for band $a$ depends only on the properties of band $a$ and the two bands represent independent channels. In any case, in the conductivity derivation below, the interband terms do not play a role, as the current-current correlation function restricts the currents to the intraband case.
\section{Interband scattering}
The spectral density function for the interband (off-diagonal) elements can be expressed in the low energy limit as
\begin{widetext}
\begin{eqnarray}
A_{\textbf{k},a,b}\left(\omega\right)&=&\frac{\Sigma'_{a,b}\left(\omega\right) \left(\Gamma_{a}\left(\omega\right)\eta_{\textbf{k},b}\left(\omega\right)+\eta_{\textbf{k},a}\left(\omega\right)\Gamma_{b}\left(\omega\right)\right)} {\left(\eta_{\textbf{k},a}\left(\omega\right)\eta_{\textbf{k},b}\left(\omega\right)- \Sigma_{a,b}'^2\left(\omega\right)\right)^2}+ \frac{\Sigma_{a,b}'\left(\omega\right)}{\eta_{\textbf{k},a}\left(\omega\right) \eta_{\textbf{k},b}\left(\omega\right)-  \Sigma_{a,b}'^2\left(\omega\right)}, \label{odsdf}
\end{eqnarray}
\end{widetext}
which clearly vanishes in both of the limits $\Sigma_{a,b}\left(\omega\right)\rightarrow 0$ (no interband scattering) and $\Sigma_{a,b}\left(\omega\right)\rightarrow \infty$ (strong interband scattering). 
\subsection{Effects of significant interband scattering on the intraband self-energy components}
\label{interbandapp}
If we avoid making the approximation that the interband elements of the self-energy are vanishingly small, we have the expression for the intraband spectral density functions Eqns. (\ref{sdf11l}), which can be rearranged to yield the expression, for $b\ne a$
\begin{widetext}
\begin{eqnarray}
A_{\textbf{k},a}\left(\omega\right) &=& \frac{-2\eta_{\textbf{k},b}^4\left(\omega\right)\left[\Gamma_{a,a}\left(\omega\right)+ \left(\frac{\Sigma_{a,b}'\left(\omega\right)}{\eta_{\textbf{k},b}\left(\omega\right)}\right)^2 \Gamma_{b,b}\left(\omega\right)\right]} {\eta_{\textbf{k},b}^4\left(\omega\right)\left[\eta_{\textbf{k},a}\left(\omega\right)- \left(\frac{\Sigma_{a,b}'\left(\omega\right)}{\eta_{\textbf{k},b}\left(\omega\right)}\right)^2 \eta_{\textbf{k},b}\left(\omega\right)\right]^2+ \eta_{\textbf{k},b}^4\left(\omega\right)\left[\Gamma_{a,a}\left(\omega\right)+ \left(\frac{\Sigma_{a,b}'\left(\omega\right)}{\eta_{\textbf{k},b}\left(\omega\right)}\right)^2 \Gamma_{b,b}\left(\omega\right)\right]^2}\nonumber\\&=&\frac{-2\left[\Gamma_{a,a}\left(\omega\right)+\left(\frac{\Sigma_{a,b}'\left(\omega\right)} {\eta_{\textbf{k},b}\left(\omega\right)}\right)^2\Gamma_{b,b}\left(\omega\right)\right]} {\left[\eta_{\textbf{k},a}\left(\omega\right)-\left(\frac{\Sigma_{a,b}'\left(\omega\right)} {\eta_{\textbf{k},b}\left(\omega\right)}\right)^2\eta_{\textbf{k},b}\left(\omega\right)\right]^2+ \left[\Gamma_{a,a}\left(\omega\right)+\left(\frac{\Sigma_{a,b}'\left(\omega\right)} {\eta_{\textbf{k},b}\left(\omega\right)}\right)^2\Gamma_{b,b}\left(\omega\right)\right]^2},\label{nzod1}
\end{eqnarray}
\end{widetext}

If we again take the low energy limit, we obtain the expressions for the spectral density function 
\begin{eqnarray}
\lim\limits_{\Gamma \rightarrow 0} A_{\textbf{k},a}\left(\omega\right) &=& -2\pi \delta\left(\eta_{\textbf{k},a}\left(\omega\right)-\left(\frac{\Sigma_{a,b}'\left(\omega\right)} {\eta_{\textbf{k},b}\left(\omega\right)}\right)^2\eta_{\textbf{k},b}\left(\omega\right)\right), \nonumber
\end{eqnarray}
and spectral density function squared
\begin{eqnarray}
\lim\limits_{\Gamma \rightarrow 0} A^2_{\textbf{k},a}\left(\omega\right) &=&\frac{-2\pi \delta\left(\eta_{\textbf{k},a}\left(\omega\right)-\left(\frac{\Sigma_{a,b}'\left(\omega\right)} {\eta_{\textbf{k},b}\left(\omega\right)}\right)^2\eta_{\textbf{k},b}\left(\omega\right)\right)} {\left[\Gamma_{a,a}\left(\omega\right)+\left(\frac{\Sigma_{a,b}'\left(\omega\right)} {\eta_{\textbf{k},b}\left(\omega\right)}\right)^2\Gamma_{b,b}\left(\omega\right)\right]}.\nonumber
\end{eqnarray}

The two band conductivity (see below for derivation of the conductivity formula) is then 
\begin{widetext}
\begin{eqnarray}
\sigma^{(0)}_{xx} &=& \hbar e^2\int\limits_{}^{}\frac{d^3\textbf{k}}{\left(2\pi\right)^3}|v_{\textbf{k},x}|^2 \int \frac{d\omega}{\left(2\pi\right)^2} \sum\limits_{a}^{} A_{\textbf{k},a}^2\left(\omega\right)\left(-\frac{dn_f(\omega)} {d\omega}\right)\nonumber\\
&=& -\hbar e^2 \int\frac{d^3\textbf{k}}{\left(2\pi\right)^3}|v_{\textbf{k},x}|^2 \int \frac{d\omega}{\left(2\pi\right)} \left[\frac{ \delta\left(\eta_{\textbf{k},1}\left(\omega\right)- \left(\frac{\Sigma_{1,2}'\left(\omega\right)}{\eta_{\textbf{k},2}\left(\omega\right)}\right)^2 \eta_{\textbf{k},2}\left(\omega\right)\right)}{\left[\Gamma_{1,1}\left(\omega\right)
+ \left(\frac{\Sigma_{1,2}'\left(\omega\right)}{\eta_{\textbf{k},2}\left(\omega\right)}\right)^2 \Gamma_{2,2}\left(\omega\right)\right]}
\right.\notag\\&&\hspace*{4.6cm}
+
\left.
\frac{ \delta\left(\eta_{\textbf{k},2}\left(\omega\right)- \left(\frac{\Sigma_{1,2}'\left(\omega\right)}{\eta_{\textbf{k},1}\left(\omega\right)}\right)^2 \eta_{\textbf{k},1}\left(\omega\right)\right)}{\left[\Gamma_{2,2}\left(\omega\right)+ \left(\frac{\Sigma_{1,2}'\left(\omega\right)}{\eta_{\textbf{k},1}\left(\omega\right)}\right)^2 \Gamma_{1,1}\left(\omega\right)\right]}\right]\left(-\frac{dn_f(\omega)} {d\omega}\right),\nonumber
\end{eqnarray}
\end{widetext}
which can be seen to approach the form given in the main text for $\left|\Sigma_{1,2}'\right|\ll\left|\eta_{\textbf{k},a}\right|$, and, in the opposite limit $\left|\Sigma_{1,2}'\right|\ll\left|\eta_{\textbf{k},a}\right|$ the corresponding delta function takes the form $\delta\left(\left(\Sigma_{1,2}'/\eta_{\textbf{k},a}\right)^2\eta_{\textbf{k},a}\right)$ and it is obvious that argument of the delta function will never vanish in this limit, and there will be no contribution to the conductivity from the $a$ band. The intermediate range is more complicated, with the limit $\left|\Sigma_{1,2}'\right|\approx\left|\eta_{\textbf{k},a}\right|$ giving an expression for the conductivity
\begin{widetext}
\begin{eqnarray}
\sigma^{(0)}_{xx}&=& \hbar e^2\int\limits_{}^{}\frac{d^3\textbf{k}}{\left(2\pi\right)^3}|v_{\textbf{k},x}|^2 \int \frac{d\omega}{\left(2\pi\right)} \left[\frac{-2 \delta\left(\eta_{\textbf{k},1}\left(\omega\right)-\eta_{\textbf{k},2}\left(\omega\right)\right)} {\left[\Gamma_{1,1}\left(\omega\right)+ \Gamma_{2,2}\left(\omega\right)\right]}\right] \left(-\frac{dn_f(\omega)}{d\omega}\right),
\end{eqnarray}
\end{widetext}
so the conductivity will vanish unless the two bands' renormalised energies overlap.
\section{Derivation of the Conductivity Formula}
In a multi-band system, the electronic current density operator component in the $\alpha$ direction $j_{\alpha}$, and the current-current correlation function $\pi_{\alpha\gamma}\left(i\omega_n\right)$ are
\begin{eqnarray}
j_\alpha&=& -e \sum\limits_{\textbf{k},\sigma,a}^{}v_{\textbf{k},\alpha,a} c_{\textbf{k},\sigma,a}^\dagger c_{\textbf{k},\sigma,a},\label{curr2b}\\
\pi_{\alpha\gamma}\left(i\omega_n\right)&=& \frac{-1}{V_D}\int\limits_{0}^{\beta}d\tau e^{i\omega_n \tau} \langle T_\tau j_\alpha \left(\tau\right)j_\gamma\left(0\right)\rangle, \label{firstcorr2b}
\end{eqnarray}
where, $V_D$ is the $D$-dimensional volume of the unit cell, $-e$ is the charge on an electron, $c_{\textbf{k},\tilde{a}}^\dagger$ and $ c_{\textbf{k},\tilde{a}}$ respectively create and annihilate an electron with momentum $\hbar\textbf{k}$ and with spin and band given by the spin-band index $\tilde{a}=\left(a,\sigma\right)$, $v_{\textbf{k},\alpha,a}= \hbar^{-1} \partial\varepsilon_{\textbf{k},\tilde{a}}/\partial k_\alpha$ is the $\alpha$-component of the (group) velocity of an electron with momentum $\hbar\textbf{k}$  in band $a$, $\tau$ is the imaginary time ($\beta=\left(k_BT\right)^{-1}$), $T_\tau$ is the imaginary time ordering operator, and $\omega_n$ the $n$th fermion Matsubara frequency, with $\langle\ldots\rangle$ denoting an ensemble average over the system. The lowest order (non-interacting) term in an $S$-matrix expansion for the correlation function is given by 
\begin{widetext}
\begin{eqnarray}
\pi^{(0)}_{\alpha\gamma}\left(i\omega_n\right)&=& \frac{-e^2}{V_D}\int\limits_{0}^{\beta}d\tau e^{i\omega_n \tau} \sum\limits_{\textbf{k},\textbf{k'};\sigma,\sigma';a,a'}^{} v_{\textbf{k},\alpha,a}v_{\textbf{k}',\gamma,a'}\langle T_\tau c_{\textbf{k},\sigma,a}^\dagger \left(\tau\right) c_{\textbf{k},\sigma,a} \left(\tau\right) c_{\textbf{k'},\sigma',a'}^\dagger \left(0\right)c_{\textbf{k'},\sigma',a'}\left(0\right)\rangle.\label{corr2b}
\end{eqnarray}

Using Wick's theorem we then rearrange the terms in the correlation function Eqn. (\ref{corr2b}), to express it in terms of non-interacting Green's functions  \cite{Mahan}
\begin{eqnarray}
\langle T_\tau c_{\textbf{k},\sigma,a}^\dagger \left(\tau\right) c_{\textbf{k},\sigma,a} \left(\tau\right) c_{\textbf{k'},\sigma',a'}^\dagger \left(0\right)c_{\textbf{k'},\sigma',a'}\left(0\right)\rangle &=&\langle c_{\textbf{k},\sigma,a} \left(\tau\right) c_{\textbf{k'},\sigma',a'}^\dagger\left(0\right)\rangle\langle c_{\textbf{k'},\sigma',a'} \left(0\right)c_{\textbf{k},\sigma,a}^\dagger \left(\tau\right)\rangle = G^{(0)}_{\tilde{a},\tilde{a}} \left(\textbf{k};\tau\right) G^{(0)}_{\tilde{a},\tilde{a}} \left(\textbf{k}; -\tau\right),\nonumber
\end{eqnarray}
where the ensemble averages enforce the conservation of momenta, as well as spin-band index. The non-interacting correlation function is then
\begin{eqnarray}
\pi^{(0)}_{\alpha\gamma}(i\omega_n) &=& \frac{e^2}{V_D} \sum\limits_{\textbf{k}.\tilde{a}}^{}  v_{\textbf{k},\alpha,\tilde{a}}v_{\textbf{k},\gamma,\tilde{a}}   \int\limits_{0}^{\beta}d\tau e^{i\omega_n  \tau}G^{(0)}_{\tilde{a},\tilde{a}} \left(\textbf{k};\tau\right) G^{(0)}_{\tilde{a},\tilde{a}} \left(\textbf{k}; -\tau\right), \label{2bmatgreen}
\end{eqnarray}
to which we introduce interactions by replacing the bare Green's functions with their corresponding dressed form \cite{Mahan}, taking interactions into account solely via the self-energy (i.e. neglecting vertex corrections). We then perform the Fourier transform to give the Matsubara frequency correlation function (from which we can find the finite temperature form via analytic continuation)
\begin{eqnarray}
\pi_{\alpha\gamma}(i\omega_n)&=& \frac{e^2}{V_D} \sum\limits_{\textbf{k},\tilde{a}}^{} v_{\textbf{k},\alpha,\tilde{a}}v_{\textbf{k},\gamma,\tilde{a}} \frac{1}{\beta} \sum\limits_{ip}^{}G_{\tilde{a},\tilde{a}} \left(\textbf{k};ip+i\omega\right) G_{\tilde{a},\tilde{a}} \left(\textbf{k}; ip\right). \nonumber\\
\label{pi0d2b}
\end{eqnarray}

The sum over Matsubara frequencies is performed by moving to the spectral representation, where the interacting Green's functions are given by  \cite{Mahan,Bruus} $G_{\tilde{a},\tilde{a}} \left(\textbf{k}; i\omega\right)=\int \frac{d\omega}{2\pi} A_{\tilde{a}}\left(\textbf{k},\omega\right)/\left(i\hbar \omega-\omega\right)$, with $A_{\tilde{a}}$ the spectral density function for an electron in spin-band $\tilde{a}$. In the limit of weak interband scattering, the interband contributions to $A_{\tilde{a}}$ vanish quadratically (see Supplementary Material for a detailed derivation), and the spectral density function is approximately that of a quasiparticle in the $\tilde{a}$ band, with only the intraband contributions contributing significantly. We then find
\begin{eqnarray}
S_{\tilde{a},\tilde{a}}(i\omega_n)&\equiv &\frac{1}{\beta} \sum\limits_{ip}^{} G_{\tilde{a},\tilde{a}} \left(\textbf{k};ip+i\omega\right) G_{\tilde{a},\tilde{a}} \left(\textbf{k}; ip\right) \nonumber\\
&=& \int \frac{d\omega_1}{2\pi} A_{\tilde{a}}\left(\textbf{k},\omega_1\right)\int \frac{d\omega_2}{2\pi} A_{\tilde{a}}\left(\textbf{k},\omega_2\right)S^{(0)}_{\tilde{a},\tilde{a}}(i\omega_n) ,\label{2bsumprd0}
\end{eqnarray}
where \cite{Mahan} 
\begin{eqnarray}
S^{(0)}_{\tilde{a},\tilde{a}}(i\omega_n)\equiv\frac{1}{\beta} \sum\limits_{ip}^{} G_{\tilde{a},\tilde{a}}^{(0)} \left(\textbf{k};ip+i\omega\right) G_{\tilde{a},\tilde{a}}^{(0)} \left(\textbf{k}; ip\right)=\frac{n_f(\omega_2)-n_f(\omega_1)} {i\hbar\omega_n+\omega_2-\omega_1}.
\end{eqnarray}
Thus,\begin{eqnarray}
\lim\limits_{\delta \rightarrow 0}\text{Im} \left\lbrace S_{\tilde{a},\tilde{a}}(\omega+i\delta)\right\rbrace &=& -\int \frac{d\omega_2}{4\pi} A_{\tilde{a}}\left(\textbf{k},\omega_2\right) A_{\tilde{a}}\left(\textbf{k},\omega_2+\omega\right)\left(n_f(\omega_2)- n_f(\omega_2+\hbar\omega)\right). \label{2bIms}
\end{eqnarray}
The conductivity is then found by taking the zero frequency limit of the retarded current-current correlation function divided by the frequency $\omega$,
\begin{eqnarray}
\sigma_{\alpha\gamma} &=&\lim_{\substack{\delta \rightarrow 0 \\ \omega\rightarrow 0}} \frac{\text{Im}\left\lbrace\pi_{\alpha\gamma}\left(\omega+i\delta\right)\right\rbrace}{\omega} \nonumber \\
&=& e^2\hbar \sum\limits_{\tilde{a}}^{} \int\limits_{-\infty}^{\infty} \frac{d^D\textbf{k}}{\left(2\pi\right)^D} v_{\textbf{k},\alpha,\tilde{a}} v_{\textbf{k},\gamma,\tilde{a}}  \int\limits_{-\infty}^{\infty} \frac{d\omega}{4\pi}  A_{\tilde{a}}^2\left(\textbf{k},\omega\right)\left(-\frac{dn_f(\omega)} {d\omega}\right). 
\end{eqnarray}
\end{widetext}

The low energy limit of the spectral density function with weak interband scattering is given by $\lim\limits_{\text{Im}\left\lbrace\Sigma\right\rbrace \rightarrow 0} A^2_{\textbf{k},\tilde{a}}\left(\omega\right) \approx -2\pi Z\delta\left(\omega-Z\varepsilon_{\textbf{k}; \tilde{a}}\right)/\Sigma''_{\tilde{a},\tilde{a}}\left(\omega\right)$, with the quasiparticle weight in band $\tilde{a}$, 
\begin{equation}
Z_{\tilde{a}}^{-1}=\left|1-\frac{\partial\Sigma'_{\tilde{a},\tilde{a}}\left(\omega\right)} {\partial\omega }\right|_{|\textbf{k}|=k_F}.
\end{equation}
The $N_b$-band conductivity is then
\begin{widetext}
\begin{eqnarray}
\sigma_{xx}&=&e^2\hbar\int\limits_{-\infty}^{\infty}\frac{d^3\textbf{k}}{\left(2\pi\right)^3} \int\limits_{-\infty}^{\infty} \frac{d\omega}{4\pi} \sum\limits_{\tilde{a}}^{2N_b} \frac{-2\pi Z_{\tilde{a}}\delta\left(\omega-Z_{\tilde{a}} \varepsilon_{\textbf{k};\tilde{a}}\right)} {\Sigma_{\tilde{a},\tilde{a}}''\left(\omega\right)}v_{\textbf{k},x,\tilde{a}}^2 \left(-\frac{dn_f(\omega)}{d\omega}\right).\nonumber
\end{eqnarray}
\end{widetext}

\bibliography{KWRRef}